\documentclass[10pt, conference, letterpaper]{IEEEtran}

\usepackage{cite}
\usepackage{amsmath,amssymb,amsfonts}
\usepackage{textcomp}
\usepackage{graphicx,xcolor,xspace}
\usepackage{algorithmic}
\usepackage[ruled,norelsize]{algorithm2e}
\usepackage{booktabs}
\usepackage{bm}

\usepackage{todonotes}

\begin{document}
\newcommand{\sysname}{eAFH\xspace}
\newcommand{\graz}{PDR-Exclusion\xspace}

\title{\sysname: Informed Exploration for Adaptive Frequency Hopping in Bluetooth Low Energy}

\author{\IEEEauthorblockN{Valentin Poirot}
\IEEEauthorblockA{\textit{Kiel University}, Germany \&\\
\textit{Chalmers University of Technology}, Sweden \\
vpo@informatik.uni-kiel.de}
\and
\IEEEauthorblockN{Olaf Landsiedel}
\IEEEauthorblockA{\textit{Kiel University}, Germany \&\\
\textit{Chalmers University of Technology}, Sweden \\
ol@informatik.uni-kiel.de}
}

\maketitle

\begin{abstract}
With more than 4 billion devices produced in 2020, Bluetooth and Bluetooth Low Energy (BLE) have become the backbone of the Internet of Things.
Bluetooth and BLE mitigate interference in the crowded 2.4~GHz band via Adaptive Frequency Hopping (AFH), spreading communication over the entire spectrum,
and further allows the exclusion of interfered channels.
However, exclusion is challenging in dynamic environments or with user mobility:
as a user moves around, interference affects new channels, forcing AFH to deprive itself of new frequencies, while some other excluded channels are now free of losses but remain excluded.
Channel re-inclusion is a primordial, yet often left out, aspect of AFH, as it is non-trivial to assess the new situation of excluded frequencies.

We present \textit{eAFH}, a mechanism for channel exclusion-and-inclusion.
eAFH introduces informed exploration to AFH: using only past measurements, eAFH assesses which frequencies we are most likely to benefit from re-including in the hopping sequence.
As a result, eAFH adapts in dynamic scenarios where interference varies over time.
We show that eAFH achieves 98-99.5\% link-layer reliability in the presence of dynamic Wifi interference with 1\% control overhead and 40\% higher channel diversity than state-of-the-art approaches.

\end{abstract}

\maketitle
\section{Introduction}
\label{eafh:sec:introduction}

Bluetooth has become the most prominent enabler of short-range wireless communications.
With more than 4 billion new devices shipped in 2020 alone~\cite{Bluetooth2021-Market}, Bluetooth is the backbone of a more pervasive IoT, with applications in smart health~\cite{Suzuki2013-Healthcare}, smart homes~\cite{Collotta2015-smarthome}, and smart cities~\cite{Spachos2020-smartcity}.
Bluetooth Low Energy (BLE), introduced in 2010, extends Bluetooth's support to devices with limited resources, and improves energy efficiency and communication range.
Yet, the rise of Bluetooth and BLE-capable equipment comes at a price: these new devices must all co-exist in an already crowded 2.4 GHz ISM band.
To survive interference caused by, e.g., Wifi communication or concurrent BLE transmissions, BLE employs Adaptive Frequency Hopping (AFH): a BLE connection periodically hops to new frequency channels.
By frequently changing channels and spreading communication over the entire spectrum, a BLE communication never stays long on the same interfered frequency and avoids consecutive losses.

However, communication will eventually hop back to interfered channels and suffer losses that could be otherwise prevented.
The BLE link layer will have to retransmit the lost packets: in low-latency and time-critical applications, frequent losses and retransmissions may induce unacceptable delays.
For example, audio traffic over Bluetooth generates 300 kbit/s, with data transmission every 15~ms.
Hopping back to interfered frequencies and suffering preventable losses inevitably degrades the user experience, as the audio quality must be decreased to handle the retransmissions and reduced bitrate, and latency increases.
To solve this, the Bluetooth standard allows devices to \textit{exclude} frequency channels from their hopping sequence~\cite{Bluetooth52}:
with channel exclusion, it becomes possible to prevent such losses.
The exclusion mechanism, i.e., how to detect bad channels, is not described by the BLE standard and left open for implementers to design their own approaches.
For example, the iOS operating system excludes Bluetooth channels overlapping with Wifi when both technologies are used on the same device, but does not exclude channels suffering external interference.
The Silicon Labs BLE stack detects external interference and excludes channels based on the received power~\cite{SiLabs-AFH}.
Sp\"ork~et~al.~show that the received power is not a good metric for channel exclusion~\cite{Spoerk2020-Timeliness}, and demonstrate that the link-layer Packet Delivery Ratio (PDR) is a better measure of channel performance and exclusion~\cite{Spoerk2020-AFH}.
Using the PDR as metric, the authors present an exclusion mechanism able to deactivate channels with degraded performance.

\textbf{Challenges.}
Channels’ performance  can  be  degraded  by  many  physical  behaviors, such as external interference and signal collisions, distance and obstacles between devices, multi-path fading~\cite{Srinivasan2008-burstiness,Kalaa2016-ble-realword,Natarajan2016-coexistence,Spoerk2020-Timeliness}.
Some failures are transient and do not repeat over time, while others, e.g., Wifi traffic, affect the channel for extended periods.
The first challenge of designing adaptive frequency hopping systems lies in the design of \textit{channel exclusion} mechanisms:
channels with degraded performance must be removed from the hopping sequence, but only if the recorded losses are not transient.
For example, concurrent BLE communication cause sporadic losses across all channels;
an exclusion mechanism should not end up excluding all channels as a result of detecting at least one collision per channel.
Further, a sufficient number of channels should be maintained in the hopping sequence, otherwise, the communication will lose the benefits of channel hopping.

Yet, a second, major challenge arises from the aspect of mobility.
As users move around carrying, e.g., smartphones and BLE peripherals, their environment characteristics evolve.
As they move closer to active Wifi access points, losses occur, and BLE channels must be excluded.
As they move away, the channels will suffer fewer losses and could be used again.
To build \textit{adaptive} frequency hopping techniques, we require the presence of \textit{channel re-inclusion} mechanisms.
However, detecting and estimating performance improvement of an excluded channel is non-trivial.
To re-include channels, we must make new measurements of their performance, using either probing or exploration.
In active probing, a central BLE device sends a control probe on an excluded channel, and a peripheral acknowledges the probe if it is successfully received.
However, probing is expensive: it requires signaling to coordinate which channel to probe and when, must not affect the timing of the established connections, and costs energy to transmit the probe and listen to the response.
Further, probing is not part of the standard: both the central and the peripheral need to implement the same mechanism to work.
Instead, we prefer to rely on exploration: excluded channels are eventually included back into the hopping sequence, thus allowing us to collect new measurements.
Exploration saves energy compared to probing: if the channel is not interfered, directly using data packets saves us the transmission of a probe.
If the channel is interfered, the link layer handles the retransmission, no data is lost, and the channel can be excluded again.
By using exploration, we can use any commercial BLE device in use today: only the central needs to implement the exploration mechanism.
Thus, the second challenge of AFH systems lies in designing efficient exploration mechanisms: only the most beneficial channels should be explored.

\textbf{Approach.}
We present \textit{\sysname}, an exclusion-and-inclusion mechanism for Adaptive Frequency Hopping.
\sysname introduces the notion of \textit{informed exploration}, a key building block to re-include channels from stale performance measurements:
\sysname combines the link-layer packet delivery ratio, nearby-channel's performance, and timeouts inspired by exponential backoffs to assess the exclusion and re-inclusion of BLE channels.
As a result, \sysname is able to adapt to dynamic scenarios in which interference varies over time, as well as re-include channels excluded due to spread-out losses from other BLE connections.
To the best of our knowledge, \sysname is the first standard-compliant system that dives into the challenge of channel re-inclusion in AFH.

\textbf{Contributions.}
This paper makes the following contributions:
\begin{enumerate}
    \setlength\itemsep{0em}
    \item We introduce the Uncertainty Measure $U$, a new heuristic allowing informed exploration when re-including channels for Adaptive Frequency Hopping;
    \item  We present \sysname, an AFH system featuring a reliability-based channel exclusion and an uncertainty-aware channel inclusion for Bluetooth Low Energy.
    \sysname only needs to be implemented by the central, and thus works with all commercial peripherals available today;
    \item We implement and distribute eAFH\footnote{Available at: redacted.} for the Zephyr RTOS and evaluate it against dynamic WiFi and BLE traffic.
    We show that eAFH achieves 98-99.5\% link-layer reliability against dynamic WiFi interference, provides a 40\% increase in channel diversity for a 1\% control overhead cost.
\end{enumerate}

\textbf{Outline.}
The paper is structured as follows:
\S\ref{eafh:sec:background} provides background on Bluetooth Low Energy and Adaptive Frequency Hopping,
\S\ref{eafh:sec:design} discusses the importance of exploration in dynamic environments, introduces the Uncertainty Measure $U$ as a driver for exploration, and presents \sysname as an uncertainty-aware AFH solution.
\S\ref{eafh:sec:evaluation} provides an evaluation of \sysname and a comparison to state-of-the-art AFH algorithms.
Finally, \S\ref{eafh:sec:related-work} summarizes the related literature and \S\ref{eafh:sec:conclusion} concludes this paper.
\section{Background}
\label{eafh:sec:background}


\subsection{Bluetooth Low Energy}
\label{eafh:sec:background:ble}

\textbf{Low Energy.}
Introduced as part of Bluetooth 4.0 in 2010, Bluetooth Low Energy (BLE) is a short-range wireless technology in the 2.4~GHz band.
It provides connectivity for platforms with low-power constraints and provides data rates ranging from 125~kb/s up to 2~Mb/s.
BLE operates on 40 2-MHz wide frequency channels, of which 37 are used for data communication, and 3 for advertisement broadcasts.
BLE mainly targets direct, one-hop communication and advertisement broadcasts, while Bluetooth Mesh extends BLE to multi-hop communication.

\textbf{Connections.}
BLE provides two operating modes: connection-less or connection-oriented communication.
In the connection-less mode, BLE devices pseudo-periodically send advertisements, that are used to establish a connection or broadcast data.
Connection-less communication only happens over the 3 dedicated advertisement channels.
The connected mode establishes communication between a pair of devices: a central device (e.g., smartphone, laptop) and a peripheral (e.g., smartwatch, headset).
To establish a connection, the central scans for advertisements from the peripheral, and replies with a connection procedure.
Connections use the 37 data channels.

Connected devices exchange data during connection events.
These events repeat every \textit{connection interval}.
An event always starts with a packet from the central to the peripheral.
Packets are then exchanged until no new data is available, or if the maximum connection duration is reached.
Note that even if no application data has to be transmitted, the central and peripheral will exchange one packet each, as keep-alive transmissions.

\textbf{Link-layer reliability.}
BLE provides reliable communication through the use of acknowledgments on the link layer.
Each packet includes a 1-bit Sequence Number (SN) and a 1-bit Next Expected Sequence Number (NESN).
A packet is acknowledged once the received NESN is not equal to the last transmitted SN.
For example, packet SN=0 is acknowledged by the peripheral if NESN=1 is received by the central.
A packet is retransmitted (within the same connection event or across connection events) until it is acknowledged.

\subsection{Adaptive Frequency Hopping in BLE}
\label{eafh:sec:background:afh}

\textbf{Frequency hopping.}
A BLE connection hops between data channels at each connection event.
A channel map $C_{map}$, unique to each connection, indicates usable and unusable frequencies out of the 37 data channels available.
Bluetooth provides two algorithms to select the next channel in the hopping sequence from the channel map: Channel Selection Algorithm (CSA) \#1 and \#2.
CSA \#1 is a low-complexity algorithm, while CSA \#2 adds the ability to change channels within a connection event, and is more secure, as its hopping sequence is harder to predict for an external observer.

\textbf{Channel map update.}
In Bluetooth 5.2, only the central (e.g., the smartphone) can modify the channel map.
Whenever channels must be added or removed from the channel map, the central initiates a Channel Map Update Procedure, and transmits the new channel map and the time at which the update will apply.
The BLE specification requires at least 6 connection events before applying the update, to ensure that the peripheral receives the update.
Thus, it induces a mandatory delay when adapting to interference, making it harder to react to sudden changes.

\textbf{Link performance.}
Different metrics can be extracted from BLE communication, such as the link-layer Packet Delivery Ratio (PDR), the per-packet received signal strength (RSSI) at the central or peripheral, or the per-channel ground noise~\cite{Spoerk2020-AFH}.
Sp\"ork~et~al.~show that the link-layer PDR is an accurate estimator of a channel quality~\cite{Spoerk2020-AFH}, and define the per-channel PDR as:
$PDR = \frac{\#ACK(P \rightarrow C)}{\#TX(C \rightarrow P)}$
where $\#ACK(P \rightarrow C)$ is the number of ACKs received by the central, and $\#TX(C \rightarrow P)$ the number of messages sent by the central.

\section{Design: Exploration and eAFH}
\label{eafh:sec:design}

\begin{figure}[tb]
    \centering
    \includegraphics[width=\columnwidth]{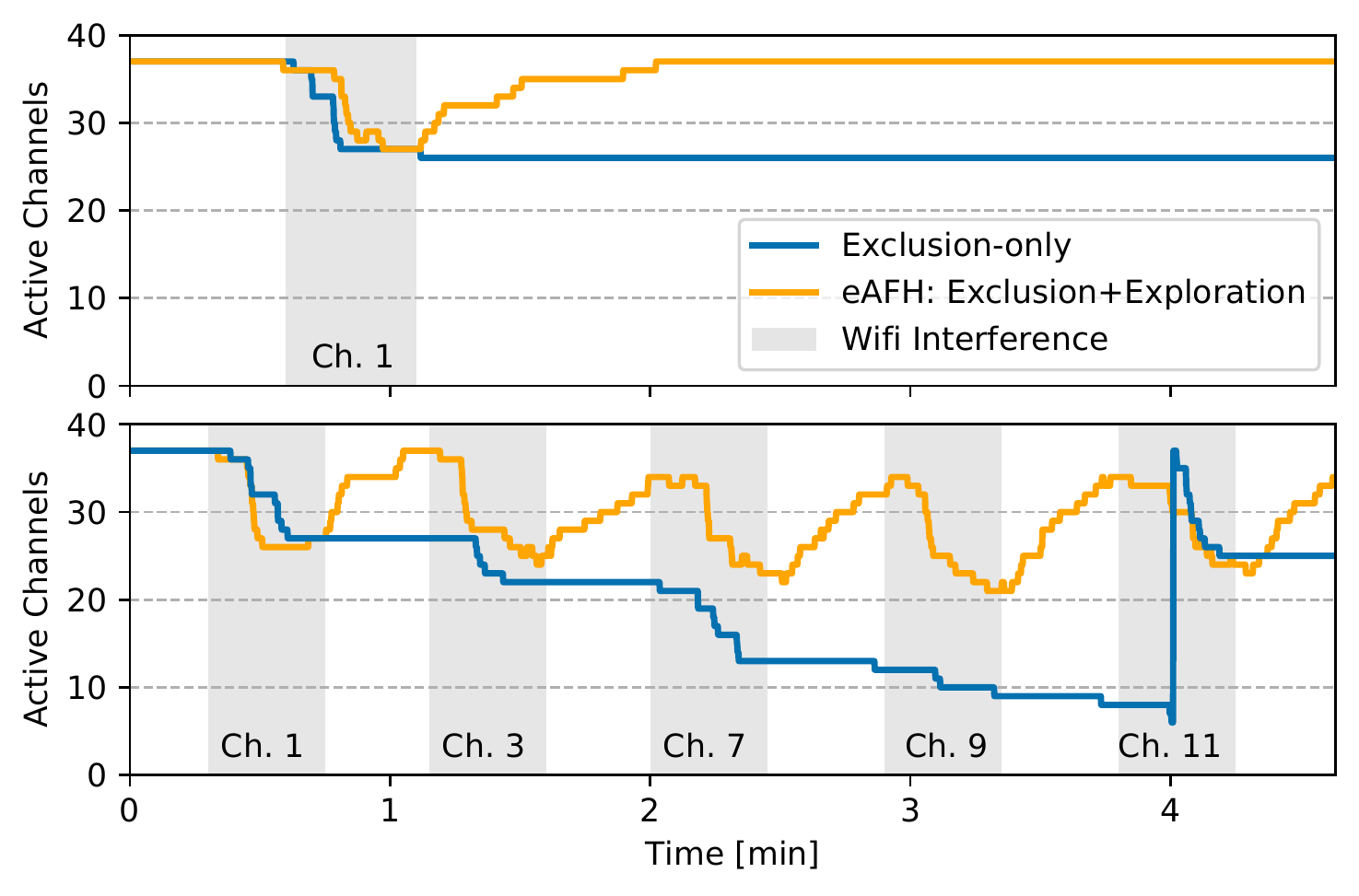}
    \caption{Importance of exploration. After suffering interference, exclusion-only systems do not re-include frequencies left out. Facing successive interference, channels are excluded until too few channels are available, after which the system resets. With exploration, eAFH maintains sufficient channel diversity at any time.}
    \label{fig:design:exploration}
\end{figure}

In this section, we begin by examining the effects of channel exclusion and exploration in the context of dynamic environments and mobile systems.
We then delve into our contributions: we introduce our uncertainty measure $U$ as a driver for informed exploration,
and present eAFH, our standard-compliant exclusion-inclusion-exploration system for Adaptive Frequency Hopping in Bluetooth Low Energy. 

\subsection{Analysis: Effects of Exclusion and Exploration}
\label{eafh:sec:design:exploration}

With its Adaptive Frequency Hopping (AFH), BLE employs constant channel hopping to avoid staying long on interfered channels, thus minimizing potential consecutive losses.
Further, BLE systems implementing channel exclusion can remove channels from their hopping sequence;
it allows exclusion-based BLE  to avoid hopping back to interfered frequencies and prevents further losses.
Interference can have many origins: other devices communicating, signals propagation and reflection, or the use of several technologies on the same platform, e.g., a smartphone featuring Wifi and Bluetooth.
For purely external interference, a channel-exclusion mechanism must find a way to model the performance of a channel, e.g., via measurements~\cite{Spoerk2020-AFH}.
In this section, we analyze how channel exclusion affects the modeling of channel performance, and how channel exploration is a key solution to build adaptive behaviors.

\textbf{Channel performance estimation.}
It is possible to design adaptive BLE systems by modeling the performance of channels, e.g., via measurements of past data transmissions~\cite{Spoerk2020-AFH}.
Using channel exclusion, such BLE systems can remove channels that have previously suffered losses from their hopping sequence.
However, as excluded channels are not used for data transmission anymore, BLE systems only relying on past data measurements cannot collect new measurements to update their performance models.
The measurements of such systems are therefore \textit{stale}: they represent a past state of the channel, but might not be representative of its current performance.
Staleness is unbounded: the last measurement might be seconds-, hours-, or even days-old.
Yet, the wireless medium is dynamic~\cite{Spoerk2020-Timeliness}: interference bursts and slow-fading affect the overall quality of channels.
In office environments, Wifi interference causes many losses during the day, yet fewer to none during the night.
Mobile devices (e.g., headphones, smartphones) will encounter many environments as the user moves around.
Exclusion is punitive here: excluded channels have potentially better performance now, but our stale measurements do not reflect their new state.

\textbf{Obtaining new measurements.}
To get new channel statistics, AFH systems must either introduce channel probing or define a scheme to re-include channels.
Probing can take two forms: passive, or active probing.
In passive probing, the central measures the noise floor of a channel: a high noise floor indicates an interfered frequency.
However, the wireless environment can differ between the central and the peripheral, and passive probing is shown to perform badly if it is executed only by the central device~\cite{Spoerk2020-Timeliness}.
Active probing is challenging: it must not impede over the connections connections maintained by the devices; it requires coordination, as the peripheral must be aware which channel will be probed, and when; and therefore requires additional signaling between the devices.
Further, probing is not part of the BLE standard, thus requiring that both the central and peripheral implement the same changes, and will not work with devices in use today.

In this paper, we propose exploration via channel re-inclusion: eventually, the central re-includes some channels back into the channel map.
These channels are thus re-used for data exchange, we obtain new measurements, and we can update the channel performance.
Channel re-inclusion is standard compliant and does not require additional implementation on the peripheral: all commercial peripherals can be used, and only the central needs to implement an adaptive re-inclusion technique.
Using data transmissions to measure the performance of a channel can be seen as risky: we are gambling with a data packet.
This is, however, not the case: BLE's link-layer provides retransmissions: a packet is retransmitted until it is acknowledged; our data is therefore never lost.
Further, probing using control packets would require at least as much energy: both the central and the peripheral are still required to send a probe, and acknowledge it, and possibly signal which and when the next channel will be probed.

\textbf{Exploring channels.}
User mobility depicts the downfall of exclusion without inclusion.
Let us picture users with a smartphone and a BLE peripheral, such as headphones, walking in a shopping mall, featuring an exclusion-only technique.
As they stroll along shops, they encounter different WiFi access points.
Although interference is localized, the BLE channels affected by WiFi traffic along the way end up excluded until almost no channels are left active.
We argue that re-including channels is necessary in dynamic environments: we propose to \textit{explore} channels by including them and collecting fresh measurements, cf. Fig.~\ref{fig:design:exploration}.
However, simply re-including channels after a static timeout is not enough: some channels will always provide low performance, while some channels may have suffered only transient disturbances earlier.
For example, in a scenario with constant Wifi interference, we should not continue to include the channels co-located with Wifi for the sake of exploration.

In this paper, we introduce \textit{informed exploration} to solve the problem of channel re-inclusion.
In a nutshell, the duration a channel is excluded for is not fixed but evolves with our knowledge of the past performance of the channel.
In \S\ref{eafh:sec:design:uncertainty}, we define an uncertainty measure to enable informed exploration, and use past performance, data staleness, and information about neighboring channels to enable efficient decisions.


\begin{figure}[tb]
    \centering
    \includegraphics[width=\columnwidth]{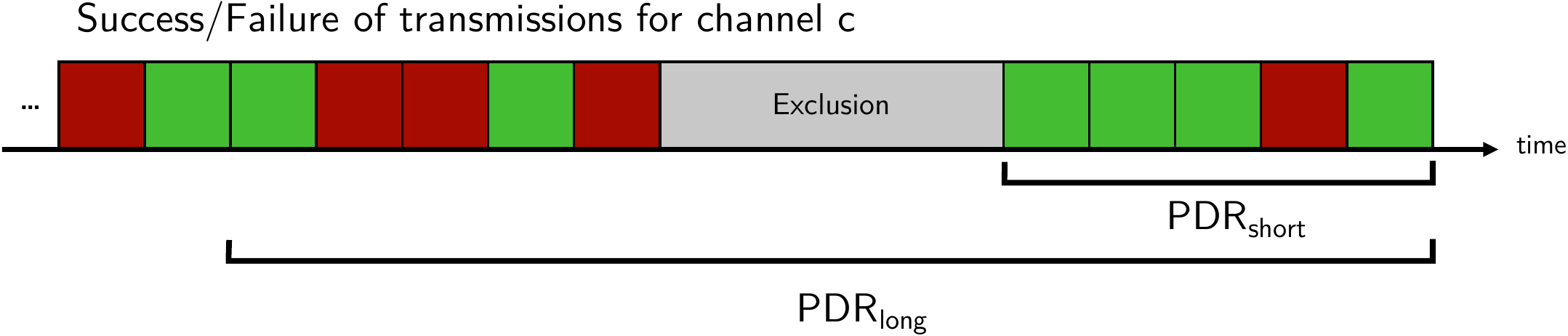}
    \caption{Measures of performance. $PDR_{short}$ (5 samples) represents the immediate performance, $PDR_{long}$ (10 samples) the average performance of a channel, respectively.}
    \label{fig:design:PDR}
\end{figure}

\begin{figure}[tb]
    \centering
    \includegraphics[width=0.8\columnwidth]{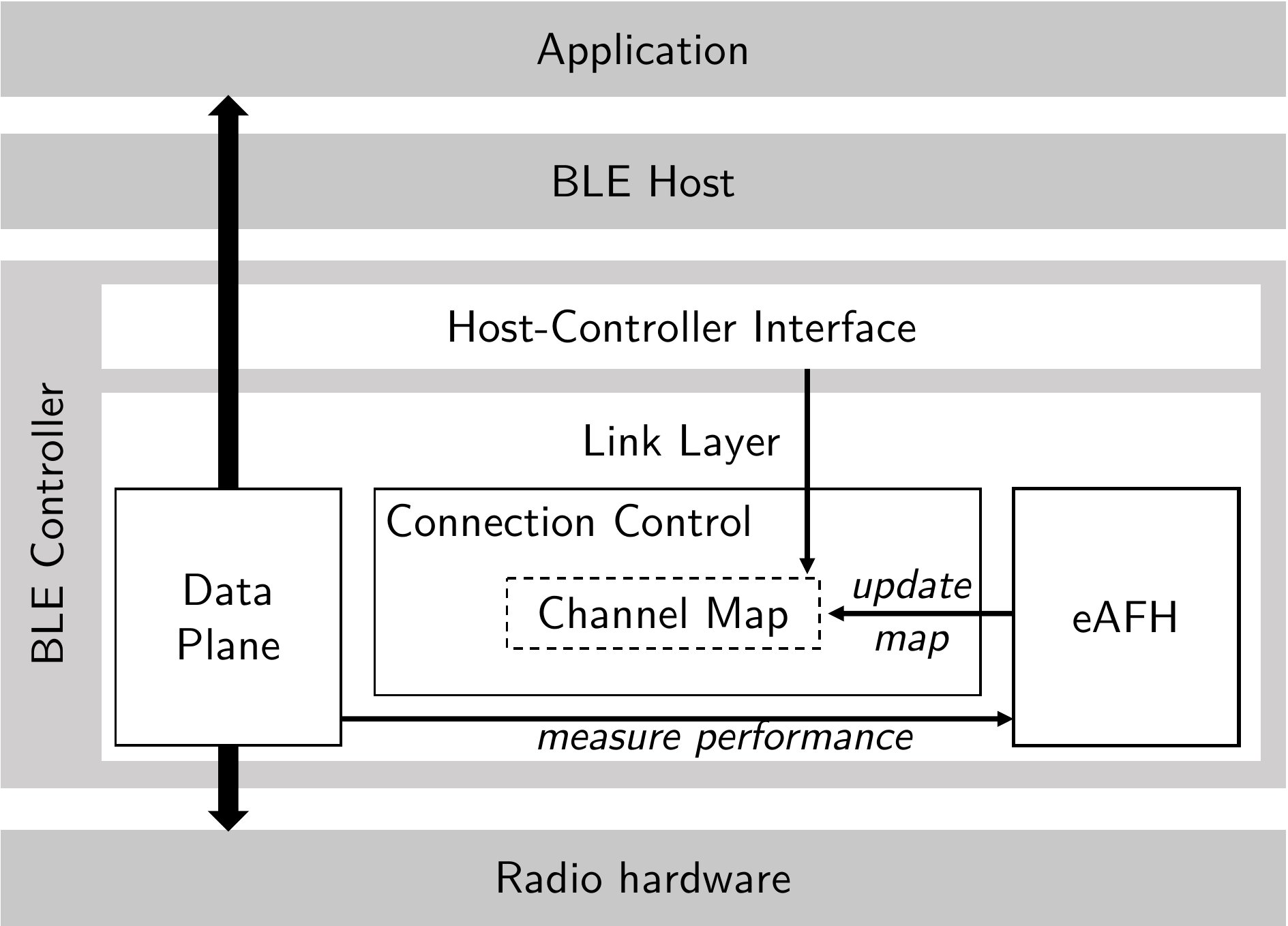}
    \caption{eAFH System Integration. eAFH collects channel performance during data transmissions, computes the channels to exclude and explore, and updates the channel map used by the connection.}
    \label{fig:design:eAFH}
\end{figure}

\subsection{Estimating Uncertainty}
\label{eafh:sec:design:uncertainty}

\textbf{Goal.}
In the absence of up-to-date channel performance, we must devise a heuristic to drive the re-inclusion of channels.
We introduce \textit{uncertainty} as metric: we quantify our uncertainty that our past measurements correctly model the current channel performance, i.e., how likely the performance has changed.
Measuring uncertainty does not require additional measurements: our confidence that past measurements model the current performance naturally decreases as time passes.
Further, we can fine-tune our uncertainty with knowledge gained from past experience: a channel suffering losses across multiple exclusions is likely to suffer from long-term interference, and it is not beneficial to explore it in the near future.
Finally, the performance of nearby channels is additional information: as Wifi interference affects multiple BLE channels, performance degradation can be correlated over nearby channels.

\textbf{Measuring performance.}
Earlier works define the link-layer packet delivery ratio $\mathit{PDR}(c)$ as the performance of the channel $c$~\cite{Spoerk2020-AFH}.
We further split the performance into two distinct metrics: the short-term and long-term performance of the channel $c$, 
$\mathit{PDR}_{short}(c)$ and $\mathit{PDR}_{long}(c)$, respectively, see Fig.~\ref{fig:design:PDR}.
$\mathit{PDR}_{short}(c)$ represents the current state of the channel $c$, while $\mathit{PDR}_{long}(c)$ represents its average performance over a longer period of time.
We compute $\mathit{PDR}_{short}$ over a sliding-window of the last transmissions and exclude statistics obtained before any channel exclusion.
In contrast, $\mathit{PDR}_{long}(c)$ keeps statistics obtained across exclusions.
As such, $\mathit{PDR}_{short}$ is an optimistic estimator: it forgets past losses, while $\mathit{PDR}_{long}$ keeps track of past losses, and is slower to adapt.
Once a channel is excluded, $\mathit{PDR}_{short}$ quickly becomes uncertain. We reset it once the channel is re-included.

\textbf{Staleness is not uncertainty}.
As channels end up excluded, their measurements become stale and their performance uncertain.
While it is easy to determine the staleness (i.e., the age) of measurements, it is not necessarily a good proxy for measuring our uncertainty of the performance $\mathit{PDR}_{short}$.
In a mobile setup, staleness is a major factor of uncertainty.
Yet, in a static setup co-located with constant Wifi traffic, data staleness is not determinant; performance is still impacted similarly by the constant interference, irrespective of the age of information.

\textbf{Uncertainty measure.}
We introduce a new heuristic, the \textit{uncertainty measure} $U$.
Our metric indicates how uncertain we are of the current performance of an excluded channel, i.e., it indicates if we should explore a channel and collect fresh performance measurements.
We construct our uncertainty measure $U$ from three indicators:
\textbf{a.} the staleness of our measurements, \textit{b.} the long-term performance of the channel, $\mathit{PDR}_{long}$; and
\textbf{c.} the performance of neighboring channels.

\textbf{Staleness and long-term performance.}
We first define the uncertainty due to data staleness, $U_{stale}$.
The older our measurements are, the more uncertain our estimated performance becomes.
We define $(t - t_{exclusion}(c))$, the time elapsed since a channel is excluded, as the data staleness.
The long-term performance is also an indicator of uncertainty: we are less likely to benefit from exploring low-performance channels, than we are from exploring higher-performance channels.
To penalize channels with lower $\mathit{PDR}_{long}$, we take inspiration from the exponential backoff mechanism: the higher the number of losses a channel has suffered, the exponentially longer it takes to explore it again.
We extract from $\mathit{PDR}_{long}$ the amount of losses we suffered,  $\mathit{losses}_{long}(c)$, and we require that an excluded channel should remain inactive for at least $D\times2^{\mathit{losses}(c)}$~seconds, where $D$ is the default exclusion duration ($D$ = 2~sec).
Finally, we can express the uncertainty due to staleness and long-term performance as:
\begin{equation}
\label{eafh:eq:U_stale}
    U_{stale}(c) = \frac{t - t_{exclusion}(c)}
                        {D\times2^{\mathit{losses}(c)}}
\end{equation}
i.e., the ratio between elapsed time since exclusion, and the required exclusion duration.
As $U_{stale}(c)$ monotonically increases, we are less certain our stale measurements correctly model the current performance of the channel.

\textbf{Correlated performance.}
Some interference affects more than one BLE channel simultaneously.
For example, Wifi transmissions overlap with multiple BLE channels, and micro-wave interference affects large parts of the spectrum.
We use this knowledge and assume that some losses are correlated between directly nearby BLE channels.
By looking at the current performance of direct neighbors, we infer if the channel is suffering from a common interference source.
We define $U_{near}(c)$ as the uncertainty caused by correlated interference, such as:
\begin{equation}
\label{eafh:eq:U_neighbor}
    U_{near}(c) = - \left( 1 - \frac{\mathit{PDR}_{short}(c-1) + \mathit{PDR}_{short}(c+1)}{2} \right)
\end{equation}
In effect, $U_{near}$ is a negative measure. It indicates that nearby channels are also affected by interference, e.g., by Wifi, and that it is not beneficial to explore this channel immediately.

\textbf{Uncertainty measure.}
We therefore define our uncertainty measure $U(c)$ such as:
\begin{equation}
\label{eafh:eq:uncertainty}
    U(c) = U_{stale}(c) +\alpha \times U_{near}(c)
\end{equation}
where $U_{stale}(c)$ and $U_{near}(c)$ are given by Eq.~\ref{eafh:eq:U_stale} and Eq.~\ref{eafh:eq:U_neighbor}, and $\alpha=2$.

\textbf{Bounded convergence.}
In our system, all excluded channels are eventually included again; it is an important feature for dynamic environments.
Thus, if our environment abruptly changes, our system eventually converges to a list of channels without excessive or outdated exclusions.

\begin{algorithm}[tb]
\scriptsize
\SetAlgoLined
\DontPrintSemicolon
\SetKwComment{Comment}{$\triangleright$\ }{}
\SetKwFunction{ChannelMapUpdateProcedure}{ChannelMapUpdateProcedure}
\KwIn{channel map $C_{map}$, last used channel $c_{t-1}$, last measurement $m$}
\KwOut{channel map $C_{map}$}
 Update $\mathit{PDR}_{short}(c_{t-1})$, $\mathit{PDR}_{long}(c_{t-1})$ with $m$\;
1. Exclusion\;
 \ForEach{channel c}
 {
  \If{$\mathit{PDR}_{short} \leq 90\%$}
    {
    $C_{map} \gets C_{map} \setminus \{c\}$\Comment*[r]{Exclude c}
    Reset $\mathit{PDR}{short}(c)$\;
    }
 }
2. Exploration\;
 \ForEach{channel c}
 {
  \uIf{$c$ in $C_{map}$}
    {
    $counter(c) \gets 0$\;
    }
    \Else{
    $counter(c) \gets counter(c) + 1$\;
    $U_{stale}(c) \gets \frac{counter(c)}{D\times 2^{losses(c)}}$\;
    $U_{nearby}(c) \gets \frac{PDR_{long}(c-1)+PDR_{long}(c+1)}{2}$\;
    $U(c) \gets U_{stale}(c) + \alpha \times U_{nearby}(c)$\;
    \If{$U(c) \geq 1.0$}
        {
        $C_{map} \gets C_{map} \cup \{c\}$\Comment*[r]{Explore c}
        }
    }
 }
3. Ensure $C_{min}$ channels\;
\If{$|C_{map}| < C_{min}$}
    {
    Include $C_{min}-|C_{map}|$ channels with the highest $PDR_{long}$\;
    }
4. Update $ C_{map}$\;
\ChannelMapUpdateProcedure($C_{map}$)\;
\Return{$C_{map}$}
 \caption{eAFH Procedure}
 \label{alg:eAFH}
\end{algorithm}

\subsection{eAFH System Integration}
\label{eafh:sec:design:system}

In this section, we present the integration of eAFH into the BLE stack.

\textbf{Controller extension.}
eAFH acts as an additional extension to the link-layer of the BLE controller and does not replace any existing mechanism, see Fig.~\ref{fig:design:eAFH}.
The central device initializes a new eAFH instance for each active connection: channel exclusion is connection-dependent, as are the channel statistics.
As peripherals may be physically far away, each connection can suffer different, localized interference.

\textbf{Execution.}
After each connection sub-event (e.g., a transmission or reception), we collect performance measurements of the current channel.
Once the connection event and all time-critical operations have been executed, we execute the eAFH algorithm, see Alg.~\ref{alg:eAFH}.
eAFH takes as input the channel map currently in use.
It executes four consecutive steps to exclude and explore channels, and to apply a new channel map to a connection.

\textbf{1. Channel exclusion.}
eAFH removes all channels where the short-term link-layer reliability $\mathit{PDR}_\mathit{short}(c)$ is insufficient.
After exclusion, we reset $\mathit{PDR}_\mathit{short}(c)$.
eAFH excludes channel $c$ from the channel map if $\mathit{PDR}_\mathit{short}(c) < T_{exclu}$, 
where $T_{exclu}=90\%$ (see evaluation in \S\ref{eafh:sec:evaluation:exclusion}). 

\textbf{2. Channel inclusion.}
We use the channel map obtained after exclusion as input to the inclusion step.
At the beginning of the inclusion, we update the staleness uncertainty (Eq.~\ref{eafh:eq:U_stale}) and neighboring channels uncertainty (Eq.~\ref{eafh:eq:U_neighbor}) for all channels.
eAFH then computes the uncertainty score such as $U(c) = U_{stale}(c) +\alpha \times U_{near}(c)$ (Eq.~\ref{eafh:eq:uncertainty}).
eAFH includes channel $c$ to the channel map if
$U(c) \geq T_{incl}$,
where the threshold $T_{incl}=1.0$, i.e., our uncertainty of this channel's performance has grown significantly.

\textbf{3. Ensuring a sufficient number of channels.}
According to the BLE standard, the channel map must contain at least $C_{min}$ active channels ($2 \leq C_{min} \leq 37$).
Only the peripheral can update this minimal number of used channels.
Following the exclusion and inclusion steps, there may be fewer than $C_{min}$ active channels.
eAFH ensures that at least $C_{min}$ channels are active by including channels with the highest long-term reliability ($ \mathit{PDR}_\mathit{long}(c)$), until the channel map contains $C_{min}$ channels.
In contrast to the short-term $\mathit{PDR}_\mathit{short}(c)$, the long-term $\mathit{PDR}_\mathit{long}(c)$ also includes measurements taken before exclusions; it represents the average, long-term performance of channel $c$.
Thus, eAFH ensures that sufficient channels are active and that active channels provide sufficient performance, based on past measurements.

\textbf{4. Channel map application.}
If eAFH desires to update the channel map, it uses the Channel Map Update Procedure of the controller's link-layer to propagate the new information to the peripheral.
The BLE standard forces that any connection update is applied after at least 6 connection events (to allow time to transmit the new parameters)~\cite{Bluetooth52}.
If a channel map update is underway, eAFH only executes steps 1 to 3.

\textbf{Implementation.}
We implement eAFH in C for the Zephyr RTOS BLE stack~\cite{ZephyrRTOS}, and make our implementation open-source (see \S\ref{eafh:sec:introduction}).
eAFH's implementation is platform-independent.
We use the nRF52840DK board as the primary target for our evaluation.
eAFH uses up to 4~kB of flash and takes 700~$\mu$sec to compute a new channel map.
Moreover, eAFH is modular and easy to extend with other channel selection algorithms.

\section{Evaluation}
\label{eafh:sec:evaluation}

In this section, we experimentally evaluate \sysname against controlled Wifi and BLE interference.
We describe our methodology and setup in \S\ref{eafh:sec:evaluation:setup}, study the exclusion mechanism in \S\ref{eafh:sec:evaluation:exclusion} and the inclusion mechanism in \S\ref{eafh:sec:evaluation:inclusion}.
Finally, we compare \sysname with state-of-the-art AFH algorithms in \S\ref{eafh:sec:evaluation:comparison}.

\subsection{Setup}
\label{eafh:sec:evaluation:setup}

We experimentally evaluate \sysname using two nRF52840DK, spaced 10~cm apart, with a 0~dBm transmit power.
Our boards are in close proximity to attenuate the impact of uncontrolled interference (e.g., nearby Wifi access-points, in-use BLE devices), while we keep the impact of our controlled interference.
We set two Raspberry Pi 4 at a distance of 20~cm away from the nRF52 boards, and use them to inject strong Wifi interference, at +20~dBm.
Note that this scenario is quite extreme, the goal being to push BLE and AFH to their limits.
We use iperf to simulate heavy TCP traffic between the jammers.

\textbf{Scenarios.}
We evaluate the following scenarios:
\begin{itemize}
    \item One-off jamming: We jam Wifi channel 1 for one minute, after which the Wifi hotspot is turned off. The medium is left unjammed for the remainder of the experiment.
    \item Continuous jamming: We jam Wifi channel 1 for the entire duration of the experiment.
    \item Fast dynamics environment: Every 30~seconds, the Wifi hotspot changes its channel, from a pseudo-random sequence containing all usable Wifi channels. As soon as the hotspot is turned on, the jammer connects and generates heavy traffic. In the slow dynamics scenario, the medium is left unjammed for 30~sec before switching the Wifi channel.
    \item BLE co-existence: We use a BLE headset connected to a laptop, and play music over it. The headset communicates roughly every 11.25~ms with the laptop.
\end{itemize}

\textbf{Metrics.}
We evaluate the following set of metrics:
\begin{itemize}
    \item Number of active channels.
    \item Link-layer Packet Delivery Ratio (PDR): the ratio of frames received (does not include retransmissions)
    \item Updates: Number of Channel Map Update Procedure calls exchanged.
\end{itemize}

\textbf{Baselines.}
We compare \sysname to the default Zephyr implementation of AFH that does not exclude any channel, that we call here \textit{No AFH}, and the state-of-the-art PDR-based channel exclusion by Sp\"ork~et~al., that we refer to as \textit{\graz}~\cite{Spoerk2020-AFH}.
In \graz, channels are excluded if their PDR drops below 95\%.
The PDR is computed over a sliding-window of 30 samples.
If less than $C_{min}$ channels are left active, \graz re-includes all channels and resets the statistics of re-included channels.

\subsection{Channel Exclusion}
\label{eafh:sec:evaluation:exclusion}

\begin{figure}[tb]
    \centering
    \includegraphics[width=\columnwidth]{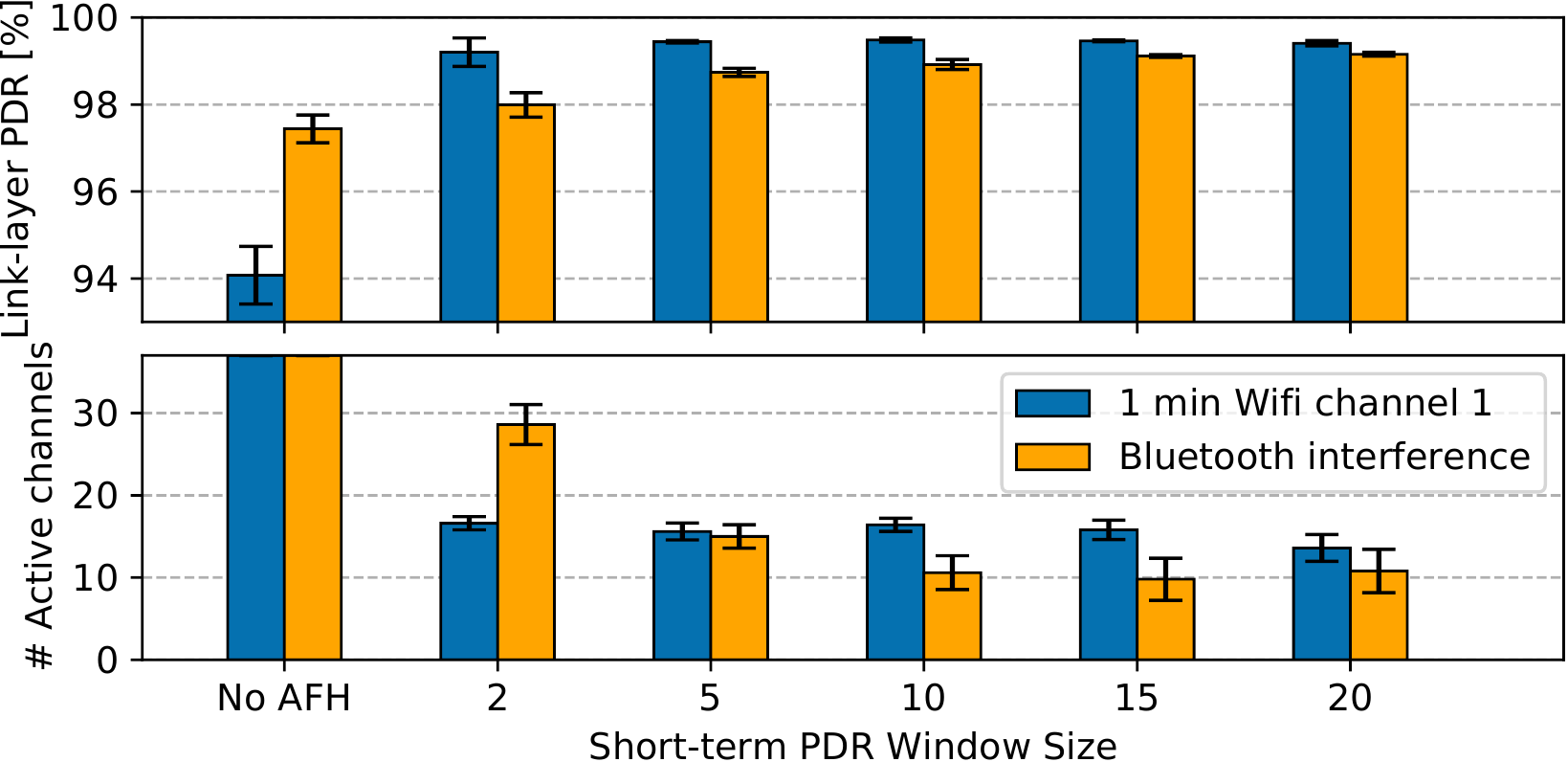}
    \caption{Effect of the sliding window size on channel exclusion. The more samples are used to compute $\mathit{PDR}_{short}$, the better reliability becomes, at the cost of many excluded channels.}
    \label{fig:eval:exclusion}
\end{figure}

\textbf{Scenario.}
We experimentally study the exclusion mechanism of \sysname by varying the number of samples required to compute $PDR_{short}$, i.e., the sliding-window size, and measure its effect on the link-layer PDR of the connection, and the number of active (non-excluded) channels.
We run two scenarios: a. 1-min one-off jamming on Wifi channel 1, and b. in the presence of BLE interference.
Channel re-inclusion is deactivated for this experiment.
We run five 5-min experiments for each value of the sliding-window size we evaluate.

\textbf{Results.}
Fig.~\ref{fig:eval:exclusion} depicts the effect of the sliding-window size on the exclusion mechanism.
The No AFH baseline, that does not contain an exclusion mechanism, provides a 94\% link-layer PDR against the strong 1-min Wifi interference.
All window sizes detect the Wifi interference and exclude at least 20 channels, up to 24 for a window-size of 20, obtaining a link-layer PDR ranging from 99.2\% to 99.4\%.
In contrast, concurrent BLE communication causes losses across the entire spectrum.
A PDR constructed with small window sizes, i.e., 2 or 5 samples, will exclude fewer channels and suffer more losses, while a PDR constructed with many measurements, i.e., 20 or more  samples, will exclude many channels, potentially excluding all channels if the experiment runs longer.
Using exclusion, the PDR is improved from 97.4\% (No AFH) to 98\% (2 samples, 8 excluded channels), up to 99.2\% (20 samples, 26 excluded channels).
For the remainder of the evaluation, we compute $PDR_{short}$ with a window size of 15, providing a good trade-off between high PDR and the number of excluded channels.

\subsection{Channel Inclusion}
\label{eafh:sec:evaluation:inclusion}

\begin{figure}[tb]
    \centering
    \includegraphics[width=\columnwidth]{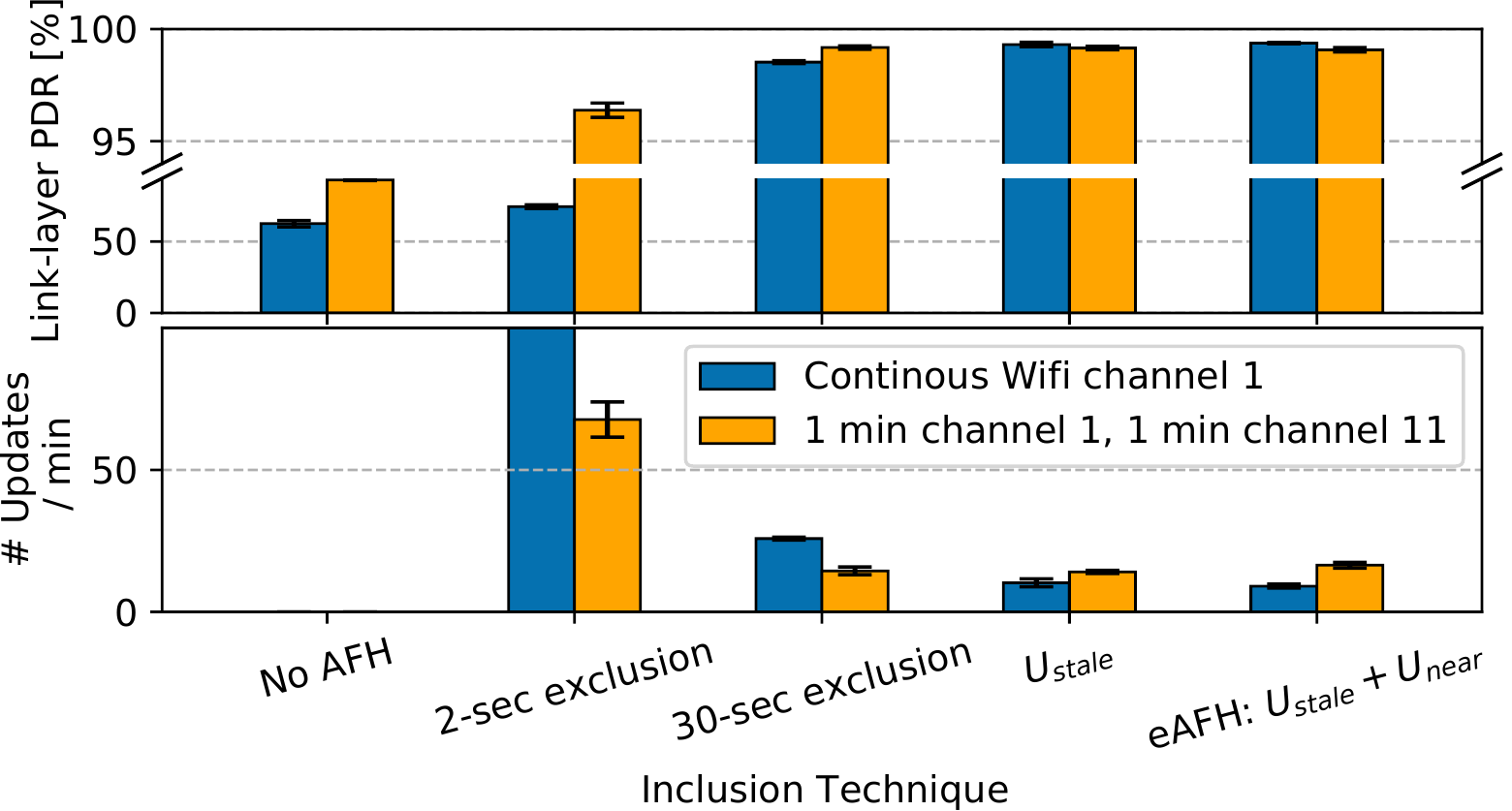}
    \caption{Channel Inclusion. Static timeouts create a large communication overhead, while \sysname's dynamic exploration provides high reliability and low overhead.}
    \label{fig:eval:inclusion}
\end{figure}

\begin{figure}[tb]
    \centering
    \includegraphics[width=0.95\columnwidth]{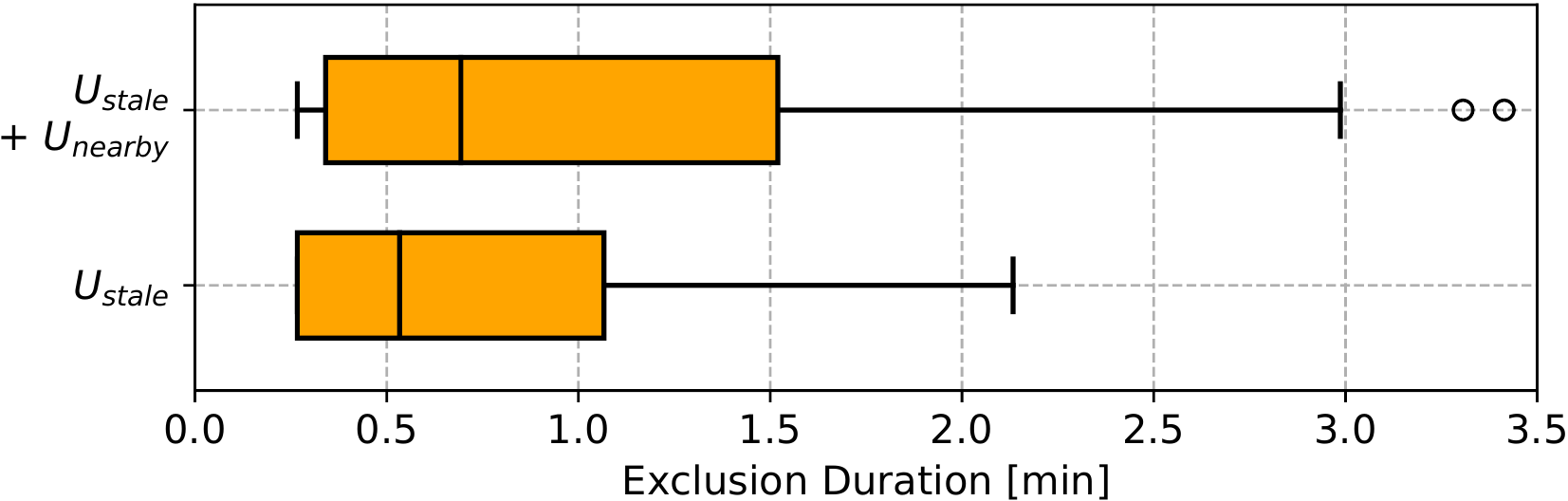}
    \caption{Improving Uncertainty using channel loss correlation. Detecting losses on nearby channels with $U_{nearby}$ leads to longer exclusion of channel interfered by Wifi signals.}
    \label{fig:eval:Unearby}
\end{figure}

\begin{figure*}[tb]
    \centering
    \includegraphics[width=\textwidth]{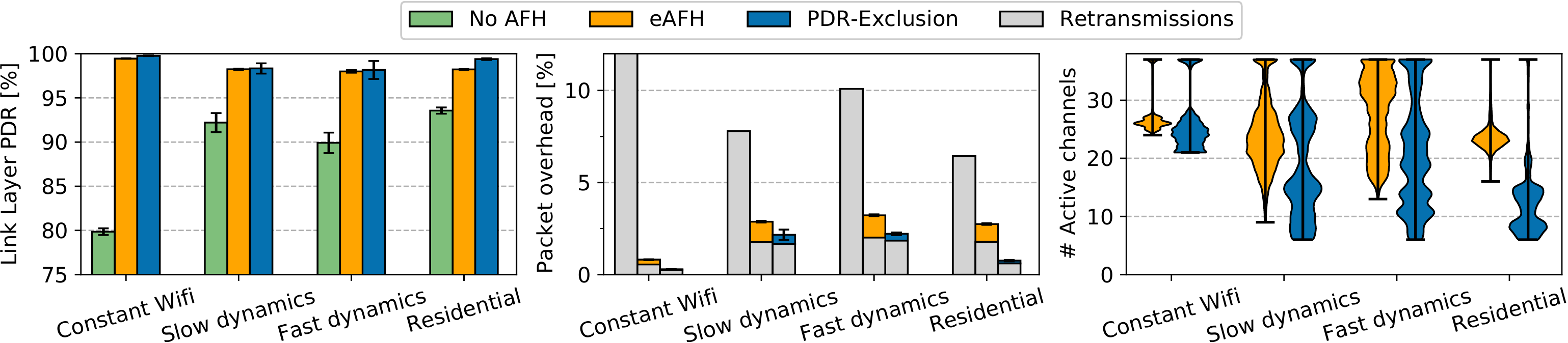}
    \caption{AFH techniques in different environments.Both \sysname and \graz greatly improve the link-layer PDR in all environments. Exploration in \sysname induces a small control overhead, but is able to maintain a higher channel diversity as a result. The AFH overhead is well below the number of retransmissions required by No AFH, meaning \sysname saves energy compared to No AFH. \graz often reaches the limit $C_{min}$ of active channels, and must reset as a result, while \sysname always provides more active channels than the limit.}
    \label{fig:eval:comparison}
\end{figure*}

\textbf{Scenario.}
We study how different inclusion techniques affect the overall performance of \sysname.
We investigate four inclusion techniques:
1. Static timeout, set to re-include a channel after 2~sec;
2. Static timeout set to 30~sec;
3. Dynamic timeout using $U_{stale}$ (see Eq.~\ref{eafh:eq:U_stale}), similar to an exponential back-off, with default inclusion after 2~sec;
4. eAFH using both $U_{stale}$ and $U_{nearby}$ (see \S\ref{eafh:sec:design:uncertainty}), inspired by exponential back-off and relying on the performance of nearby channels.
We evaluate these approaches in two scenarios:
a. continuous jamming, where frequent exploration is detrimental, and
b. one-off jamming, where frequent exploration is beneficial.
We measure the overall link-layer PDR of the BLE connection with each inclusion method, and the number of channel map updates exchanged between the central and the peripheral, i.e., how many control packets are required by exploring channels.
We run five 5-min experiments for each approach, for each scenario.

\textbf{Results.}
Fig.~\ref{fig:eval:inclusion} depicts how channel inclusion affects BLE communication.
Against continuous jamming, the No AFH baseline only achieves 62\% link-layer PDR.
Re-including channels after 2 sec increases reliability to 74\%, but induces 257 channel map updates per minute, as we need to include and exclude interfered channels every two seconds.
As we communicate every 20~msec, or over 3000 packets per minute, this represents an 8.6\% communication overhead.
By increasing the timeout, we can reduce the overhead due to channel exploration.
The 30-sec timeout achieves 98.5\% PDR and requires only 25 updates per minute, i.e., a 0.86\% overhead.
By using exponential backoff, \sysname requires fewer updates as time passes.
The $U_{stale}$-only approach achieves 99.3\% and requires 10.2 updates per minute (0.34\% overhead), while \sysname using both $U_{stale}$ and $U_{nearby}$ achieves 99.4\% with 9.1 updates per minute (0.30\% overhead).
This demonstrates the importance of exponential-based timeouts against long-term interference.

In the presence of two short, 1-min Wifi bursts,
the No AFH baseline achieves 93\% PDR.
The 2-sec timeout improves reliability and achieves 96.4\% PDR, by requiring 68 updates per minute (2.3\% overhead).
The 30-sec timeout is theoretically close to the best inclusion here, as it explores only once during the interference.
It achieves 99.2\% PDR with 14 updates per minute (0.5\% overhead).
In contrast, \sysname starts with short, 2-sec exclusion before exploring interfered channels and increasing its exclusion timeout.
Although it explores more than the 30-sec timeout, $U_{stale}$ obtains similar performance: 99.2\% PDR, with 14 updates per minute (0.5\% overhead).
\sysname achieves 99.1\% PDR with 16 updates per minute (0.5\% overhead).
These results show, although theoretically at disadvantage against short bursts, \sysname achieves similar results as constant timeouts closely matching the interference duration while outperforming them against long-term interference.

\textbf{Effect of \textit{U\textsubscript{nearby}}.}
The inclusion evaluation demonstrates that exponential backoff $U_{stale}$ accounts for an important part in the performance of \sysname.
We now investigate how $U_{nearby}$ can improve the performance of \sysname.
The role of $U_{nearby}$ is to extend the exclusion duration for channels if nearby channels are interfered.
We run \sysname with and without $U_{nearby}$, against continuous jamming on Wifi channel 1, for 5 minutes, and repeat the experiment 10 times.
Fig~\ref{fig:eval:Unearby} depicts the distribution of how long channels interfered by Wifi are excluded.
When eAFH uses both $U_{stale}$ and $U_{nearby}$, channels are excluded 37\% longer than when eAFH uses $U_{stale}$ as its sole driver for exploration.
In contrast, channels that are excluded but did not suffer Wifi interference, i.e., excluded channels that do not overlap with Wifi channel 1, were not excluded longer using $U_{nearby}$ (no difference at all).
These results demonstrate that using losses correlated across nearby channels, \sysname can avoid exploring channels that are affected by large-band interference such as Wifi.

\begin{figure}[tb]
    \centering
    \includegraphics[width=1.0\columnwidth]{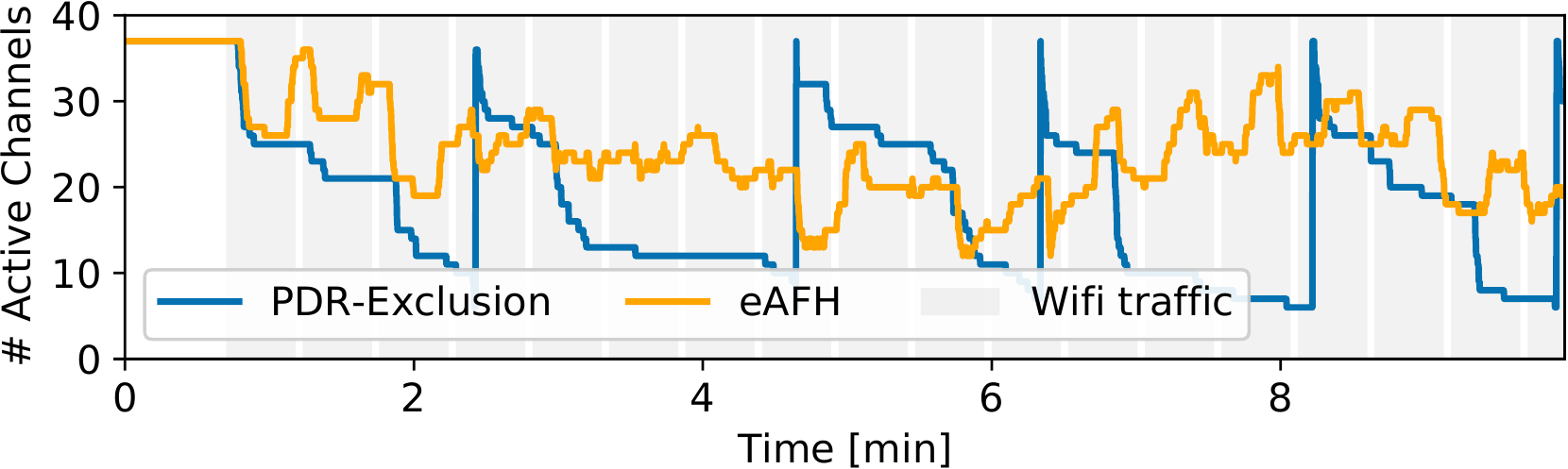}
    \caption{Representative run of \sysname (PDR: 98.1\%) and \graz (96.7\%) against Wifi interference hopping between channels. \graz continuously excludes channels until the minimum is reached, and resets its channel map.}
    \label{fig:eval:macro_vs_fast_dynamics}
\end{figure}

\subsection{State-of-the-art Comparison}
\label{eafh:sec:evaluation:comparison}
 
\textbf{Scenario.}
We now investigate the performance of \sysname against two baselines:  No AFH: the default AFH implementation of the Zephyr RTOS, \graz: the PDR-driven channel exclusion proposed by Sp\"ork et al.~\cite{Spoerk2020-AFH}.
\graz provides the best link-layer reliability in the literature, and re-includes channels once too many channels are excluded from the channel map.
We study the behavior of all three approaches in four scenarios:
a. against constant Wifi interference;
b. against slowly changing Wifi interference;
c. against highly dynamic Wifi interference;
d. against real-life disturbances in a residential building;
We measure the link-layer reliability, communication overhead due to channel map updates and number of active channels.
In the residential building scenario, the two BLE devices are standing two meters apart, while they stand in close proximity in all former scenarios.
In the slow dynamics scenario, we generate Wifi traffic on a pseudo-randomly chosen Wifi channel for 30~sec, leave the medium free of disturbances for 30~additional seconds, before selecting a new Wifi channel and repeating the traffic.
In the fast dynamics scenario, we do not leave the medium free: every 30~sec, we select a new Wifi channel and generate TCP traffic immediately.
The slow and fast dynamics scenarios represent mobile environments, where users move in the vicinity of different access-points.

\textbf{Results.}
Fig.~\ref{fig:eval:comparison} presents the link-layer PDR, communication overhead, and the distribution of the number of active channel maps of all approaches.
In all scenarios, \sysname and \graz greatly improve reliability compared to the No AFH baseline by excluding channels.
Against constant Wifi traffic, the baseline No AFH achieves 79.9\% link-layer PDR, i.e., 20\% of all packets need to be retransmitted at least once due to losses, while \sysname and \graz achieve 99.5\% and 99.8\%, respectively.

In dynamic environments, the No AFH baseline achieves 92.2\% and 89.9\% reliability against slow and fast dynamics respectively.
Both \sysname and \graz provides improved performance, 98.2\% and 98\% for \sysname, 98.3\% and 98.2\% for \graz.
The exploration mechanism of \sysname has a slightly larger communication overhead; \sysname induces a 1.1\% communication overhead to explore, while \graz creates a 0.5\% overhead by excluding channels.
In the slow dynamics scenario, No AFH requires an 8\% communication overhead to retransmit failed packets,
\sysname requires 2.9\% overhead: 1.8\% for retransmissions, and 1.1\% for exploration.
Thus, exploration using \sysname remains more efficient than keeping all channels active.
\graz requires a 2.2\% total overhead in the slow dynamics scenario.

Furthermore, \sysname always improves channel diversity compared to \graz: as different Wifi bands are active, \graz excludes more and more BLE channels, until only $C_{min}$ channels are left in the channel map, at which point \graz resets all channels, as depicted in Fig.~\ref{fig:design:exploration} and Fig.~\ref{fig:eval:macro_vs_fast_dynamics} for slow and fast dynamics, respectively.
\sysname provides a median number of 23 active channels against slow dynamics, and 29 channels against fast dynamics, while \graz provides a median number of active channels of 16 against slow dynamics, and 19 against fast dynamics.
More importantly, \sysname always provides more than the $C_{min}$ active channels required, and never reaches the $C_{min}$ limit.
In contrast, \graz reaches $C_{min}$ on average 3.2 times, and thus resets 3.2 times in 10 minutes, under fast dynamics.
Under slow dynamics, \graz resets 2.6 times.
In a residential setting with devices set 2 meters apart, \sysname achieves 98.2\% reliability with 24 active channels, while \graz achieves 99.2\% by deactivating most channels.

\textbf{Main findings.}
\sysname achieves 98-99.5\% link-layer PDR against static and dynamic interference.
The 1.1\% communication overhead due to exploration remains smaller than the 6-20\% overhead due to retransmissions in the No AFH case, i.e., \sysname saves energy.
Finally, with its exploration, \sysname provides 40\% more active channels than \graz in dynamic settings, for relatively close reliability improvements.

\section{Related Work}
\label{eafh:sec:related-work}

\textbf{Adapting BLE.}
Sp\"ork et al.~evaluate the selection of metrics for channel exclusion in BLE\cite{Spoerk2020-AFH}.
They demonstrate that the channel PDR is the best estimator of a channel's performance, as the noise floor and Signal to Noise Ratio measurements do not encompass all possible communication failures.
The authors present a channel exclusion mechanism building on the PDR as estimator.
However, they do not present a re-inclusion scheme.
The channel map and all measurements reset once too many channels are excluded.
In contrast, we introduce informed exploration as a heuristic and build an inclusion mechanism into eAFH.
We also further split the PDR into two estimators, $PDR_{short}$ and $PDR_{long}$, to drive our exclusion and re-inclusion techniques.
Sp\"ork et al.~also evaluate how adapting the physical modes of BLE improves performance in the presence of weak signal strength, i.e., when devices are far apart, but cannot improve performance in the presence of interference, as channel exclusion does.

Other works propose to modify the Channel Selection Algorithm (CSA) used by BLE to better avoid interfered channels.
Pang et al.~introduce the Interference Awareness Scheme (IAS)~\cite{Pang2021-IAS}: a channel is excluded if a connection-event timeout is triggered, i.e., no packets could be received during a connection event, or if too many packets are lost during a connection event.
In addition, the authors propose a new CSA: the channels are weighted by the probability that interference affects the channel.
The hopping sequence then skips channels that are likely to be interfered.
Cheikh et al.~ introduce SAFH, and also rely on weights to improve channel selection~\cite{Cheikh2011-SAFH}.
Channels with a low Frame Error Rate have a lower probability to be used next, or excluded altogether, while good channels are more likely to be used.
In eAFH, we do not modify the CSA.
Instead, we explore excluded channels via re-inclusion to monitor their current performance.
As a result, \sysname only needs to be implemented by the central and works with all commercial peripherals, while both IAS and SAFH require modified BLE peripherals to operate.

Further, BLE provides additional parameters that can be tuned to react to changes in the environment.
AdaptaBLE controls the transmission power, BLE physical mode and connection interval to optimize the energy consumption and Quality of Service of BLE communication~\cite{Park2020-AdaptaBLE}.
AdaptaBLE and \sysname are two orthogonal approaches, and can be used together to further improve Quality of Service and reliability.
Yang and Tseng propose channel partitioning for BLE extended advertisements~\cite{Yang2020-AdvBlacklist}.
BLEX improves connection event scheduling in systems where one central connects to multiple peripherals~\cite{Park2021-BLEX}.

\textbf{Frequency Hopping in wireless technologies.}
Multiple wireless technologies feature frequency hopping to mitigate interference, such as IEEE 802.15.4e-TSCH~\cite{Watteyne2015-TSCH,Du2012-ATSCH,Elsts2017-adaptive-TSCH}, Bluetooth Classic~\cite{Lee2009-BluetoothFH,RS-AFH,Golmie2003-AFHsched}, 5G~\cite{Li2017-5G}, and to access unlicensed bands~\cite{Popovski2006-unlicensed}.
In TSCH, A-TSCH measures and excludes channels with high background noise via RSSI measurements~\cite{Du2012-ATSCH};
Elsts et al.~combine RSSI measurements and the Packet Reception Rate (PRR) to detect channels with low performance~\cite{Elsts2017-adaptive-TSCH}.
LABeL uses an exponential weighted moving average estimator of the PDR to model the performance of a 6TiSCH channel~\cite{Kotsiou2017-LABeL}.
Machine learning techniques have also been explored: MABO-TSCH uses Multi-Armed Bandits to estimate link quality~\cite{Gomes2015-TSCHBlacklist}, while
Farahmand and Nabi use self-supervised deep learning to model future link quality and drive blacklisting~\cite{Farahmand2021-selfsupervisedTSCH}.
Carhacioglu et al.~propose to jointly schedule 802.15.4 and BLE communication on dual-radio devices to cooperatively avoid interference~\cite{Carhacioglu2018-coexistence}.
Few works investigate blacklisting the 80 channels available in Bluetooth Classic:
Lee et al.~propose to use adjacent channels' PER~\cite{Lee2009-BluetoothFH}, while Rhode-Schwarz use packet losses~\cite{RS-AFH}.
\section{Conclusion}
\label{eafh:sec:conclusion}

In Bluetooth Low Energy (BLE), the Adaptive Frequency Hopping (AFH) mechanism is primordial in mitigating external interference.
By constantly hopping between frequency channels, AFH systems avoid consecutive losses, and by excluding channels from the hopping sequence, exclusion-enabled AFH systems avoid hopping back to interfered channels.
We present \textit{\sysname}, a system with exclusion and inclusion mechanisms for Adaptive Frequency Hopping.
\sysname introduces the notion of \textit{informed exploration} as a key building block to re-include channels from stale performance measurements.
As a result, \sysname is able to adapt in dynamic scenarios where interference varies over time, as well as re-include channels excluded due to spread-out losses from other BLE connections.
In \sysname, only the central needs to implement the exploration mechanism, \sysname thus works with all commercial peripherals available today. 
We experimentally show that eAFH achieves 98-99.5\% link-layer reliability in the presence of dynamic Wifi interference, improves channel diversity by 40\% compared to state-of-the-art approaches, while only inducing a 1\% control overhead.


\bibliographystyle{IEEEtran}
\bibliography{references}

\end{document}